\documentclass[a4paper,fleqn,usenatbib]{mnras}

\usepackage[T1]{fontenc}
\usepackage{ae,aecompl}

\usepackage{graphicx}	
\usepackage{amsmath}	
\usepackage{amssymb}	

\title[Residual velocities of SMC star clusters]{Residual velocities of Small Magellanic Cloud star clusters}

\author[Andr\'es E. Piatti]{
Andr\'es E. Piatti$^{1,2}$\thanks{E-mail: andres.piatti@unc.edu.ar} \\
$^{1}$Instituto Interdisciplinario de Ciencias B\'asicas (ICB), CONICET-UNCUYO, Padre J. Contreras 1300, M5502JMA, Mendoza, Argentina\\
$^{2}$Consejo Nacional de Investigaciones Cient\'{\i}ficas y T\'ecnicas, Godoy Cruz 2290, C1425FQB,  Buenos Aires, Argentina\\
}

\date{Accepted XXX. Received YYY; in original form ZZZ}

\pubyear{2021}

\begin{document}
\label{firstpage}
\pagerange{\pageref{firstpage}--\pageref{lastpage}}
\maketitle

\begin{abstract}
We analyzed the largest Small Magellanic Cloud (SMC) cluster sample
(32) with proper motions and radial velocity measurements, from which we obtained
their space velocity components. By adopting as a reference the recent best-fitted 
rotating disc of SMC star clusters as a function of the position angle, we computed
their residual velocity vectors, and compared
their magnitudes ($\Delta$$V$) with that of a cluster with residual velocity 
components equal to the velocity dispersions along the three independent SMC
rotating disc axes of motion ($\Delta$$V$ = 60 km/s). We found that clusters that 
belong to SMC tidally induced structures have $\Delta$$V$ $>$ 50 km/s, which
suggests that space velocities of clusters in the process of escaping the  rotating
disc kinematics, are measurably different. Studied clusters pertaining to a northern 
branch of the Magellanic Bridge, the main Magellanic Bridge, the Counter-Bridge 
and the West halo give support to these findings. NGC\,121, the oldest known
SMC cluster, does not belong to any SMC tidal feature, and has $\Delta$$V$ = 
64 km/s, slightly above the boundary between bound and kinematically
 perturbed clusters. 
 \end{abstract} 




\section{Introduction}

Recently, \citet{piatti2021b}  obtained a rotating disc for 
Small Magellanic Cloud (SMC) star clusters using available radial velocities, 
derived proper motions, and positions of 25 clusters with ages between $\sim$ 1 
and 10 Gyr old. He computed the three rotational velocity components 
\citep{vdmareletal2002} of each cluster for 5.3$\times$10$^8$ possible combinations 
of values of right ascension and declination of its centre, of heliocentric distance of 
its centre, of radial velocity and proper motions in right ascension and declination, of 
inclination of the disc, of position angle of the line-of-nodes, of rotational velocity, 
and of velocity dispersions along the three independent axes of motion.
The resulting  rotating disc, represented by a combination of parameter values that
best resembled the observed clusters' movements, turned out to be very
similar to that best-fitted by \citet{zivicketal2020} for field SMC giants 
\citep[see, also][]{niederhoferetal2021,deleoetal2020,gaiaetal2020b}. This means
that both galaxy components are kinematically synchronized at some level for
distances smaller than $\sim$ 2 kpc from the SMC centre.

\newpage

The SMC clusters' motions derived by \citet{piatti2021b} show also some
noticeable scatter around the obtained representative rotating disc, which
is not caused by the considerable range of the clusters' ages used. 
Instead, \citet{piatti2021b} found that the velocity dispersion plays an important role
in dealing with such a scatter around the mean rotation curve. Particularly,
he found that velocity dispersions of 50, 20, and 25 km/s along the right
ascension, declination, and line-of-sight axes, respectively, reasonably
represent the observed velocity dispersion. This finding reveals that a rotating
disc is not enough to describe the SMC clusters' kinematics and therefore,
there exists clusters with motions that depart from that of  the rotation on a plane
around the SMC centre.  According to \citet{piatti2021b}, the  rotating disc is 
roughly  a tilted edge-on disc  (disc inclination = (70.0$\pm$10.0)$\degr$, 
position angle of the line-of-nodes = (200.0$\pm$30.0)$\degr$), with a stretching along the direction perpendicular
to it, that nearly coincides with that of the Magellanic Clouds connecting bridge.

It has been shown that the SMC is under tidal effects by the interaction with the
Large Magellanic Cloud \citep[LMC,][]{pieresetal2017,zivicketal2018,mackeyetal2018,deleoetal2020,niederhoferetal2021,omkumaretal2021}. \citet{beslaetal2012} showed that the observed irregular morphology 
and internal kinematics of the Magellanic System (in gas and stars) are naturally 
explained by interactions between the LMC and SMC, rather than gravitational 
interactions with the Milky Way. They examined the gas and stellar kinematic centres 
of the LMC; the warped LMC old stellar disc and bar; the gaseous arms stripped out 
of the LMC by the SMC in the direction of the Magellanic Bridge; the stellar debris 
from the SMC seen in the LMC disc field; etc., to strongly reinforce the suspicions of 
\citet{dVF1972} that the interaction with the Milky Way is not responsible for the 
LMC/SMC’s morphology. Moreover, these conclusions provide further support that the Magellanic Clouds are completing their first infall to the Milky Way. Hence, a question arises unavoidably: Are SMC tidal structures
kinematically different from that of the SMC rotating disc? or in other words, do 
kinematically perturbed clusters have any kinematic signatures that differentiate them from 
bound clusters? \citet{diasetal2021} observed clusters located in a branch of the
Magellanic Bridge, placed toward the northeastern outskirts of the SMC; in the SMC
main body; and in the so-called Counter-Brigde, that is  nearly superimpose in the sky 
to that  Magellanic Bridge branch. They estimated heliocentric distances, radial 
velocities, and proper motions for a sample of 7 clusters that can be useful, once
they are converted into space velocity components, to probe the existence of any 
distinguishable kinematic features between clusters belonging to these three different
galaxy structures, two of them (Magellanic Bridge and Counter-Bridge) originated by 
tidal interaction  with the LMC \citep{db2012,beslaetal2016}. 

In this work, we show that there exist different space residual velocities  for
clusters that are spatially distributed throughout SMC tidal structures 
and those bound to the SMC main body. This outcome allowed us to propose
a simple kinematic criterion to assess at first glance whether a SMC cluster is
bound to the galaxy or belong to any  kinematically perturbed stellar component. In Section
2 we describe the gathered data employed in this work, while in Section
3 we analyze and discuss the kinematic characteristics of SMC clusters distributed
at projected distances in the sky between $\sim$ 1 kpc and 4 kpc from the SMC 
centre.

\section{Data gathering}

We used the $v_1$, $v_2$, and $v_3$ space velocity components according to
\citet{vdmareletal2002} of 25 SMC clusters analyzed by \citet{piatti2021b} and
those of the 7 clusters studied by \citet{diasetal2021}. As far as we are aware,
this is the largest sample of SMC clusters with derived space velocity components.
They were calculated from measured radial velocities and proper motions, and the
transformation equations (9), (13), and (21) in 
\citet{vdmareletal2002}\citep[see, also][]{piattietal2019}. We started from the
hypothesis of, if a cluster has been perturbed by the LMC, its circular motion should 
change, and in the process of escaping from the SMC, its velocity behavior 
will be measurably different. In order to avoid projection effects of the clusters' motions
space velocity components are suitable; inspecting proper motions or radial
velocities could be misleading.

\begin{figure}
\includegraphics[width=\columnwidth]{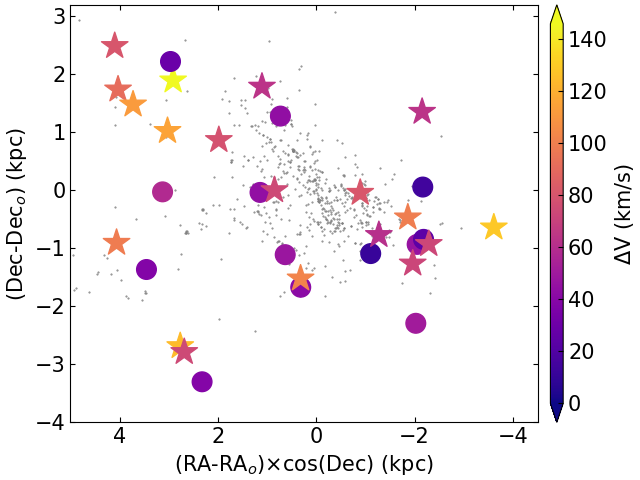}
\caption{Spatial distribution of SMC clusters compiled in \citet{bicaetal2020} 
(gray dots).  Coloured circles and stars correspond to clusters with 
residual velocities smaller and larger than 60 km/s.}
\label{fig1}
\end{figure}

We also searched the literature for accurate heliocentric distances of the
cluster sample. Unfortunately, only the 7 clusters studied by \citet{diasetal2021}
(B\,168, BS\,188, BS\,196, HW\,56, HW\,85, IC\,1708, L1) and 12 clusters
analyzed by \citet{piatti2021b} (K\,1, K\,3, K\,4, K\,8, K\,44, L\,1, L\,110, L\,113, 
NGC\,121, NGC\,339, NGC\,361, NGC\,419) have available distances. Their
values and corresponding uncertainties were taken from \citet{cetal01,glattetal2008a,diasetal2016,paetal14,mvetal2021}; and 
\citet{diasetal2021}. We also computed the position angles of the clusters
from north to east.  Table~\ref{fig1} lists the values of position angles and
distances adopted, alongside with their respective references.

From a careful reading of the references used to compile SMC clusters' distances, it 
turns out to be that the estimation of distances of SMC clusters has not been
targeted by any observational campaign yet. On the contrary, there is a very limited 
number of SMC clusters with distance measurements. Most of them have been obtained 
from Padova's isochrone fitting using very sophisticated statistical methods 
\citep[see, e.g.][]{diasetal2016,diasetal2021} and/or considering spectroscopic 
metallicities \citep[see, e.g.][]{glattetal2008a,paetal14,diasetal2021}.
The compiled distances are the most valuable treasure available up-to-date to carry out 
the analysis described in Section\,3, whose results must be interpreted as those
based  on the available cluster sample. The outcomes found in 
this work are pioneers to uncover an SMC picture that allows us to question about the 
nature and consequences of the interactions between both Magellanic Clouds. Precisely, 
this pioneering approach  anticipates further studies that will be hopefully carried out from observational  data obtained from the new generation telescopes.

Table~\ref{tab:tab1} indicates the method employed to estimate the adopted distances, namely:
theoretical isochrone fits to the cluster colour-magnitude diagrams, magnitudes of the cluster red giant clumps, 
or the Luminosity-Period relationship for $\delta$ Sct stars. We adopted here as a reference distance
scale that of the Padova theoretical isochrones \citep{betal12}, because $\sim$ 65$\%$ of the cluster
sample have distance estimates that rely on those isochrones. Using 7 clusters in our sample with distances
estimated from fitting of Teramo \citep{pietrinfernietal2004} and Dartmouth \citep{dotteretal2007} isochrones, we found
a mean difference in the distance modulus of (Teramo - Dartmouth) of 0.009$\pm$0.021 mag, and for
4 clusters in common in our sample with distances from Padova and Teramo isochrone fits, a difference in the distance modulus of -0.008$\pm$0.020 mag.
 We note that a difference of 0.01 mag in the distance
modulus implies a distance difference of 0.27 kpc at a distance of 60 kpc. Therefore, we assumed that
there is no offsets between distances obtained from these sets of theoretical isochrones that could affect
 the subsequent analysis. As for distances derived from estimation of the magnitude of the Red Clump, 
 we found 4 clusters \citep[K\,3, L\,1, NGC\,121, 339;][]{cetal01} with distances estimated using the Red Clump 
 method and  Padova isochrone fits (see Table~\ref{tab:tab1}) and the mean difference turned our to be 
 (Red Clump - Padova iso) = -2.5$\pm$1.2 kpc. Therefore, we corrected the distances of K\,44, L\,113 and
 NGC\,361 by adding 2.5 kpc to the mean values listed in Table~\ref{tab:tab1}. As for the distance of
 NGC\,419, we found a mean difference of 0.1 kpc with respect to the value estimated from Padova
 isochrone fits \citep{glattetal2008a}.

\begin{table*}
\caption{Properties of SMC star clusters.}
\label{tab:tab1}
\begin{tabular}{@{}lccccc}\hline
ID & PA (deg) & $\Delta$$V$ (km/s) & $D$ (kpc)  & Method & Ref.\\\hline
Bruck168 & 50.8	& 29.2$\pm$49.4	&61.9$\pm$2.1	 &Padova iso	&6\\
BS188 	&67.9	& 115.7$\pm$40.5	& 52.7$\pm$3.0 &Padova iso	&6\\
BS196 	&56.9	& 80.6$\pm$35.7	& 50.1$\pm$2.0 &Padova iso	&6\\
HW56 	&30.3	& 63.8$\pm$40.0	& 53.5$\pm$1.2 &Padova iso	&6\\
HW85 	&65.6	& 112.1$\pm$44.0	& 54.0$\pm$1.6 &Padova iso	&6\\
IC1708 	&54.0	& 145.5$\pm$48.5	& 65.2$\pm$1.5 &Padova iso	&6\\
K1		&247.2	& 73.1$\pm$34.2	& 53.7$\pm$2.4 &Teramo \& Dartmouth iso	&4\\
K3 		&270.9	& 14.6$\pm$27.0	& 60.6$\pm$1.1 &Padova, Teramo \& Dartmouth iso	&2\\
K4 		&248.0	& 24.4$\pm$37.2	& 54.9$\pm$2.3 &Teramo \& Dartmouth iso	&4\\
K8 		&255.0	& 99.2$\pm$24.0	& 69.8$\pm$2.3 &Padova iso	&3\\
K44 		&149.5	& 48.0$\pm$32.2	& 62.2$\pm$2.7 &Red Clump	&1\\
L1 		&259.5	& 129.2$\pm$20.7	& 56.9$\pm$1.0 &Padova, Teramo \& Dartmouth iso	&2\\
L100		&61.4	& 78.1$\pm$36.3	& 58.6$\pm$0.7 &Padova iso 	&6\\
L110 	&90.8	& 58.2$\pm$32.7	& 47.9$\pm$2.3 &Teramo \& Dartmouth	&4\\
L113 	&102.7	& 98.8$\pm$33.2	&50.5$\pm$1.7	 & Red Clump	&1\\
NGC121 	&302.2	& 63.6$\pm$25.9	& 64.9$\pm$1.2 &Padova, Teramo \& Dartmouth iso	&2\\
NGC339 	&168.9	& 42.0$\pm$29.7	& 57.6$\pm$4.1 &Padova, Teramo \& Dartmouth iso	&2\\
NGC361 	&31.3	& 44.2$\pm$16.9	& 55.8$\pm$1.7 &Red Clump	&1\\
NGC419 	&92.7	& 44.6$\pm$21.2	& 56.2$\pm$1.3 &$\delta$ Sct L-P relation	&5\\
\hline
\end{tabular}

\noindent Ref.: (1) \citet{cetal01}; (2) \citet{glattetal2008a}; 
(3) \citet{diasetal2016};  
(4) \citet{paetal14};  \\
(5) \citet{mvetal2021}; (6) \citet{diasetal2021}. 
\end{table*}

\section{Analysis and discussion}

The rotating disc geometry obtained by \citet[][see his Fig. 4]{piatti2021b} can be 
described as a 3D vector with components $v_1$, $v_2$, and $v_3$ that
are functions of the position angle (PA). Therefore, the differences between each 
velocity component of a cluster and the respective ones for the rotational curve 
at the cluster's PA ($\Delta$$v_i$, i=1,2,3) represent the components of the 
residual velocity vector with  respect to the SMC  rotating disc. By adding them
in quadrature we obtained the magnitude of such a vector $\Delta$$V$ =
($\Sigma$$_{i=1}^{i=3}$ $\Delta$$v_i$$^2$)$^{1/2}$. Figure\,1 shows the
spatial distribution of the 32 clusters with calculated $v_1$, $v_2$, and $v_3$
velocity components by adopting the parameters of the disc obtained
by \citet[][see his Table\,2]{piatti2021b}. As can be seen, clusters span a wide
range of  $\Delta$$V$  values, that we refer as the residual velocity vector.

By using as residual velocity vector
components the values of the dispersion velocity found by
\citet{piatti2021b} (i.e., $\Delta$$v_1$=25 km/s, $\Delta$$v_2$=50 km/s, 
$\Delta$$v_3$=20 km/s), we derived the
corresponding residual velocity vector magnitude, $\Delta$$V$ = 60 km/s. 
We adopted this value to differentiate clusters bound to the SMC
rotating disc ($\Delta$$V$ $<$ 60 km/s) from those scattered by the tidal
interaction with the LMC ($\Delta$$V$ $>$ 60 km/s). With 
this  residual velocity cut, we found 13 clusters with a kinematics that very 
well match that of a  rotating disc, and others 19 whose orbital motions we 
interpreted have been affected by the effects of tidal forces. Figure~\ref{fig1}
depicts their spatial distribution. Clusters with $\Delta$$V$ $>$ 60 km/s are mainly 
located in SMC regions projected in the sky that have some previous hints for 
tidal structures arisen from the interaction between both Magellanic Clouds, 
namely: a northern branch of the Magellanic Bridge and the Counter-Bridge,
which are superimposed at the northeastern outskirts of the SMC; the main
Magellanic Bridge, located toward the east from the SMC centre; the West
Halo \citep{db2012,diasetal2016} placed at the southwestern end of the SMC disc.
This finding confirms our suspicions that  kinematically perturbed clusters -- those with orbital 
motions different from those on a  rotating disc -- accelerated by the LMC 
gravitational field. Likewise, there are some clusters with $\Delta$$V$ $>$ 60 
km/s that appear projected toward the SMC main body. As far as we are
aware, this is the first observational evidence of the distinctive hot kinematics
of clusters that depart from the SMC cold disc kinematics. Note that the 
rotation velocity of the SMC disc is 25.0$\pm$5.0 km/s \citep{piatti2021b,zivicketal2020}.

Two additional features can be glimpsed at Fig.~\ref{fig1}. On the one hand,
some clusters with $\Delta$$V$ $<$ 60 km/s are also projected on
the above mentioned SMC tidal structures, which could imply that
SMC disc clusters may be projected on galaxy tidal features, in very
good agreement with the disc extension out to  $\sim$ 4 kpc 
\citep[see Fig. 1 in][]{diasetal2021}. Therefore, the sole projected cluster position 
is not enough to straightforwardly identify any clusters as
being tidally  perturbed.  We do not rule out the possible existence of cold 
kinematics clusters in the SMC tidal structures. On the other hand, some clusters 
with $\Delta$$V$ $>$ 60 km/s are seen projected on the SMC main body, which 
could witness the presence of runaway clusters traveling throughout the SMC disc. 

In order to disentangle whether there is a possible mixture of  kinematically perturbed
clusters 
with $\Delta$$V$ smaller and larger than 60 km/s, which would reveal that
they would not be easily recognized from their  residual velocity 
vectors, their heliocentric distances are necessary. Unfortunately, there is a small
number of clusters with measured distances (see Section 2), which points to the 
need of  observational campaigns in order to build a realistic spatial distribution 
of them \citep{piatti2021a}, and consequently, to trace more precisely the 3D 
SMC cluster population structure. Previous star cluster studies adopted, in general, 
a mean 
SMC distance when dealing with cluster colour-magnitudes diagrams
\citep[][and references therein]{p12b,p14c,petal2016}. This is
because changes in the distance modulus by an amount equivalent to the average 
depth of the SMC \citep{cetal01,ripepietal2017,graczyketal2020}, so that clusters can be 
placed in front or behind the SMC, leads to a smaller age difference than that 
resulting from the  isochrones bracketing the observed star cluster features in the 
colour-magnitude diagram. 

\begin{figure}
\includegraphics[width=\columnwidth]{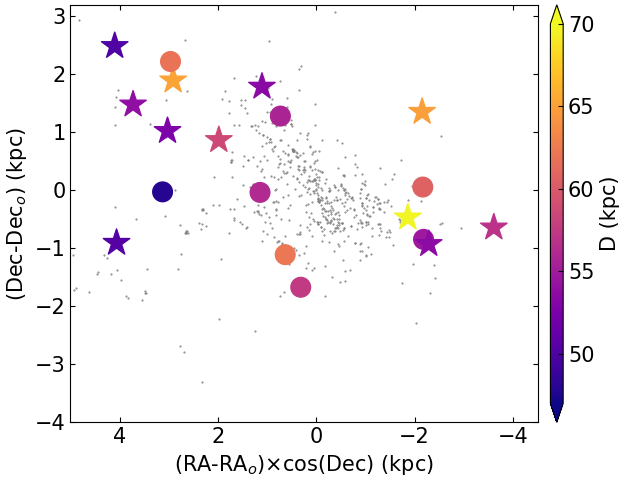}
\caption{Spatial distribution of SMC clusters compiled in \citet{bicaetal2020} 
(gray dots). Circles and stars correspond to clusters with 
 residual velocities smaller and larger than 60 km/s, colour-coded
according to their heliocentric distances.}
\label{fig2}
\end{figure}

With available heliocentric distances for 19 clusters
in our sample, we built Fig.~\ref{fig2}, which confirms that there is a wide
cluster distance range among those projected on SMC tidal structures,
in comparison with clusters projected toward the SMC main body. In the
northeastern outermost region, there are clusters located in from of the SMC 
(they belong to the northern branch of the Magellanic Bridge); in the SMC body 
itself; and behind the galaxy (they are part of the Counter-Bridge)
\citep{diasetal2021}. These clusters have very different $\Delta$$V$ values as 
judged by the different symbols used to represent them (circles and stars as in
Fig.~\ref{fig1}). Similarly, the line-of-sight toward the West Halo also
presents clusters placed in front of the SMC; at the SMC distance; and 
behind it. This is the first evidence from clusters's distances that the West
Halo is a perturbed structure of the SMC, consisting of  leading and trailing tails
\citep{db2012,diasetal2016}. Nevertheless,  in order to determine the extension 
of both tails, further distance estimates for a larger sample of  clusters projected 
on this sky region are desirable. We also showed that SMC disc clusters 
($\Delta$$V$ $<$ 60 km/s) are spread throughout the whole covered area and 
located at distances compatible with them being part of the SMC main body 
\citep[56 $\la$ $D$ (kpc) $\la$ 62,][and references therein]{piatti2021b}.

\begin{figure*}
\includegraphics[width=\textwidth]{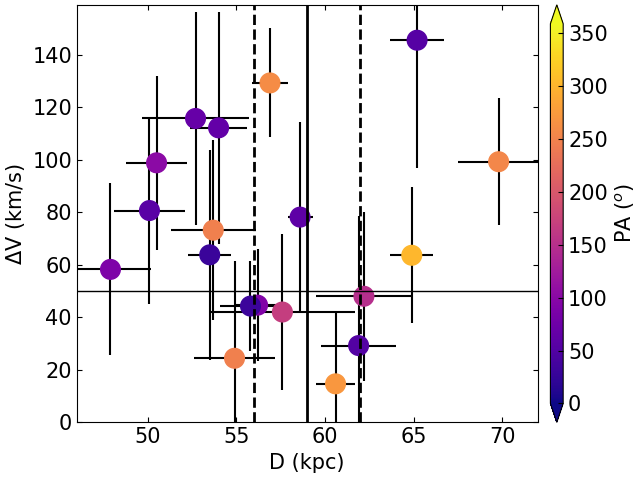}
\caption{Residual velocities $\Delta$$V$ versus heliocentric distance
of SMC clusters. The uncertainties in $\Delta$$V$ were calculated using
the measured errors in proper motion and radial velocity, and those from the 
solution for the 3D movement of the SMC centre obtained by \citet{piatti2021b}, 
propagated through the transformation equations and added in quadrature. 
Solid and dashed lines represent the SMC centre obtained by \citet{piatti2021b} 
and a readily visible SMC main body depth, respectively.}
\label{fig3}
\end{figure*}

Figure~\ref{fig3} reveals a striking distribution of SMC clusters bound
to the SMC main body and those pertaining to SMC tidal structures. According to
the observed distances of the Magellanic Bridge 
\citep[$D$  $<$ 55 kpc,][]{wagnerkaiser17,jacyszyndobrzenieckaetal2020}; of the
Counter-Bridge \citep[$D$ $>$ 65 kpc,][]{diasetal2021}; and the present West
Halo depth (52 $\la$ $D$ (kpc) $\la$ 70) derived from 5 clusters at PA between 230$\degr$ 
and 270$\degr$, there is a suggestive  residual velocity level at $\sim$ 50 km/s
that broadly separates  SMC main body clusters from those in tidally induced
features. Clusters that belong to these last structures show  residual velocities
$\Delta$$V$ $>$ 50km/s, in very good agreement with the  residual velocity 
cut calculated above (60 km/s) from the SMC dispersion velocities obtained by
\citet{piatti2021b}. SMC main body clusters span a wide range of
residual velocities, but most of them show $\Delta$$V$ $<$ 50 km/s, meaning
that they have motions similar to the SMC  rotating disc kinematics \citep{zivicketal2020}.
This outcome shows that the kinematics of SMC tidal
structures differentiates from that of its disc. 

Among clusters with heliocentric distances smaller than 55 kpc, there are
four that belong to the northern branch of the Magellanic Bridge (PA $<$ 90$\degr$);
another one belongs to the main branch of the Magellanic Bridge (PA $\approx$ 
1$\degr$); and one is part of the leading West Halo. All these clusters have 
$\Delta$$V$ $>$ 50. km/s. As clusters placed behind the SMC ($D$ $>$ 65 kpc)
are considered, there is one that pertains to the Counter-Bridge, and another 
to the trailing West Halo. They also have $\Delta$$V$ $>$ 50. km/s. 
NGC\,121 is the oldest known SMC cluster \citep[10.5 Gyr,][]{p11b}; its location does 
not coincide with any tidally induced structure 
(((RA-RA$_0$)$\times$cos(Dec),(Dec-Dec$_0$)) $\approx$ (-2.0 kpc, 1.3 kpc)),
see Fig.~\ref{fig2}), but the cluster is at $D$ = (64.9$\pm$1.2) kpc and has
$\Delta$$V$ = 64 km/s. It is within the kinematic regime of tidally perturbed clusters.
The lack of ancient globular clusters in the SMC has been one of the most intriguing enigmas in our understanding of the SMC formation and evolution. \citet{carpinteroetal2013} modeled the dynamical LMC-SMC interaction and their
respective star cluster populations aiming at exploring whether the lack of
old SMC globular clusters can be the result of such galaxy tidal interactions.
They found that clusters that originally belonged to the SMC are more likely to be
found in the outskirts of the LMC or ejected into the intergalactic medium.
In this context, NGC\,121 could be identified as one of those SMC old escaping clusters.

There are two clusters located at distances compatible with the SMC main body
depth whose  residual velocity vectors ($\Delta$$V$ $>$ 50 km/s) suggest that
they do not belong to the SMC  rotating disc. These clusters are projected
toward the northern branch of the Magellanic Cloud and the West Halo, 
respectively, so that they could either indicate that at the very onset of both tidal 
structures clusters' kinematics change or that they are runaways clusters.
We note that other runaway clusters have been identified in the LMC \citep{piattietal2018a}.
The remaining eight clusters with distance estimates in our sample are SMC
main body objects with an SMC  rotating disc kinematics ($\Delta$$V$ $<$ 50 km/s).

\begin{figure*}
\includegraphics[width=\textwidth]{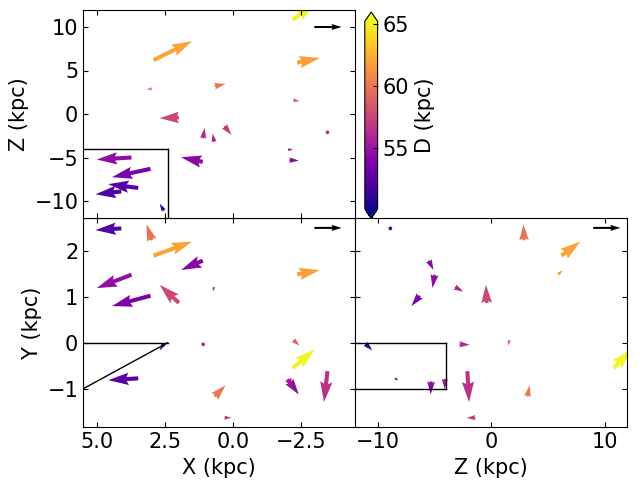}
\caption{Projected spatial distributions of SMC clusters, with the corresponding
velocity vectors. The black arrow  represents a velocity of 1$\pm$km/s.
The projection of the Magellanic Bridge is also schematically
represented by black lines \citep{wagnerkaiser17,jacyszyndobrzenieckaetal2020,bicaetal2020}. }  
\label{fig4}
\end{figure*}

For completeness purposes, Fig.~\ref{fig4} illustrates the residual velocity 
components projected on the Cartesians planes \citep{vdmareletal2002}.
They witness the stretching experienced by the star cluster population due to
tidal effects: those on the northern branch of the Magellanic Bridge roughly moving 
toward the  LMC, while the one observed in the Counter-Bridge traveling in an
oposite direction; a picture that is also verified among clusters placed in the closer 
and farther regions of the West Halo.

 In what follows, we mention two different scenarios that could be considered
to speculate on possible interpretations of the nature and consequences of
the tidal interaction between the LMC and the SMC. They are not exhaustive, but
simply introductory, based on recent results.

The gaps and spurs seen in the spatial distribution of Magellanic Bridge's clusters
\citep[see Fig. 3 in][]{bicaetal2020} resemble those observed in tidal tails or streams
around Milky Way globular clusters 
\citep[e.g., Pal\,5, GD-1 streams:][]{gd2006a,gd2006b}, which have become the 
most sensitive tracers of the nature and distribution of dark matter in the
Milky Way \citep{bonacaetal2019}. 
These tidal tails resulted to be exceedingly cold, 
with stellar velocity dispersions on the order of 2 km/s or less, and mean motions
similar to their parent globular clusters \citep{pricewhelanetal2019}. Moreover, the
discovery of tidal tails around Milky Way globular clusters using astrometric
information usually relies on the assumption that stars in tidal tails share the mean
proper motions and radial velocities of globular clusters' stars \citep[e.g.,][]{carballobello2019,sollima2020,shippetal2020}.
The presence of tidal stream substructures extending to hundreds of parsecs around 
globular clusters also suggest that at least some globular clusters continue to be 
embedded in dark matter halos \citep{olszewskietal2009,kuzmaetal2016}.
\citet{diteodoroetal2019} studied high-resolution H\,I data in the SMC and found that
out to $\sim$ 4 kpc from its centre, the galaxy exhibits a rotating disc kinematics
that possibly flattens outwards, despite of the strong gravitational interaction with 
the LMC. In order to reproduce the observed rotation curved, they argued that
a dominant dark matter halo is required. Recently, \citet{koposovetal2018} 
discovered a mildly elliptical ultra-faint dwarf galaxy (Hydrus) located in the 
Magellanic Bridge with a velocity dispersion of 2.7$\pm$0.5 km/s, which indicates 
that the system is dark matter dominated. 

To the light of our knowledge about the kinematics of globular clusters' tidal
tails, the Magellanic Bridge and West Halo do not behave similarly with
respect to the SMC. We found that they show motions
largely departed from that of the SMC main body (see Fig~\ref{fig3}).
Therefore, the present observational evidence would be supporting either the 
notion of a tidal interaction pattern between the LMC and the SMC not comparable 
to that known for globular clusters and the Milky Way, or that the SMC
is not a dark matter dominated dwarf galaxy. We think that Fig.~\ref{fig3} is
a suitable starting point to be considered for further investigations on such a 
debatable conundrum.

\section{Data availability}

Data used in this work are available upon request to the author.

\section*{Acknowledgements}
I thank the referee for the thorough reading of the manuscript and
timely suggestions to improve it. 




\begin{thebibliography}{}
\makeatletter
\relax
\def\mn@urlcharsother{\let\do\@makeother \do\$\do\&\do\#\do\^\do\_\do\%\do\~}
\def\mn@doi{\begingroup\mn@urlcharsother \@ifnextchar [ {\mn@doi@}
  {\mn@doi@[]}}
\def\mn@doi@[#1]#2{\def\@tempa{#1}\ifx\@tempa\@empty \href
  {http://dx.doi.org/#2} {doi:#2}\else \href {http://dx.doi.org/#2} {#1}\fi
  \endgroup}
\def\mn@eprint#1#2{\mn@eprint@#1:#2::\@nil}
\def\mn@eprint@arXiv#1{\href {http://arxiv.org/abs/#1} {{\tt arXiv:#1}}}
\def\mn@eprint@dblp#1{\href {http://dblp.uni-trier.de/rec/bibtex/#1.xml}
  {dblp:#1}}
\def\mn@eprint@#1:#2:#3:#4\@nil{\def\@tempa {#1}\def\@tempb {#2}\def\@tempc
  {#3}\ifx \@tempc \@empty \let \@tempc \@tempb \let \@tempb \@tempa \fi \ifx
  \@tempb \@empty \def\@tempb {arXiv}\fi \@ifundefined
  {mn@eprint@\@tempb}{\@tempb:\@tempc}{\expandafter \expandafter \csname
  mn@eprint@\@tempb\endcsname \expandafter{\@tempc}}}

\bibitem[\protect\citeauthoryear{{Besla}, {Kallivayalil}, {Hernquist}, {van der
  Marel}, {Cox}  \& {Kere{\v s}}}{{Besla} et~al.}{2012}]{beslaetal2012}
{Besla} G.,  {Kallivayalil} N.,  {Hernquist} L.,  {van der Marel} R.~P.,  {Cox}
  T.~J.,   {Kere{\v s}} D.,  2012, \mn@doi [\mnras]
  {10.1111/j.1365-2966.2012.20466.x}, \href
  {http://adsabs.harvard.edu/abs/2012MNRAS.421.2109B} {421, 2109}

\bibitem[\protect\citeauthoryear{{Besla}, {Mart{\'{\i}}nez-Delgado}, {van der
  Marel}, {Beletsky}, {Seibert}, {Schlafly}, {Grebel}  \& {Neyer}}{{Besla}
  et~al.}{2016}]{beslaetal2016}
{Besla} G.,  {Mart{\'{\i}}nez-Delgado} D.,  {van der Marel} R.~P.,  {Beletsky}
  Y.,  {Seibert} M.,  {Schlafly} E.~F.,  {Grebel} E.~K.,   {Neyer} F.,  2016,
  \mn@doi [\apj] {10.3847/0004-637X/825/1/20}, \href
  {http://adsabs.harvard.edu/abs/2016ApJ...825...20B} {825, 20}

\bibitem[\protect\citeauthoryear{{Bica}, {Westera}, {Kerber}, {Dias}, {Maia},
  {Santos}, {Barbuy}  \& {Oliveira}}{{Bica} et~al.}{2020}]{bicaetal2020}
{Bica} E.,  {Westera} P.,  {Kerber} L. d.~O.,  {Dias} B.,  {Maia} F.,  {Santos}
  Jo{\~a}o F.~C. J.,  {Barbuy} B.,   {Oliveira} R. A.~P.,  2020, \mn@doi [\aj]
  {10.3847/1538-3881/ab6595}, \href
  {https://ui.adsabs.harvard.edu/abs/2020AJ....159...82B} {159, 82}

\bibitem[\protect\citeauthoryear{{Bonaca}, {Hogg}, {Price-Whelan}  \&
  {Conroy}}{{Bonaca} et~al.}{2019}]{bonacaetal2019}
{Bonaca} A.,  {Hogg} D.~W.,  {Price-Whelan} A.~M.,   {Conroy} C.,  2019,
  \mn@doi [\apj] {10.3847/1538-4357/ab2873}, \href
  {https://ui.adsabs.harvard.edu/abs/2019ApJ...880...38B} {880, 38}

\bibitem[\protect\citeauthoryear{{Bressan}, {Marigo}, {Girardi}, {Salasnich},
  {Dal Cero}, {Rubele}  \& {Nanni}}{{Bressan} et~al.}{2012}]{betal12}
{Bressan} A.,  {Marigo} P.,  {Girardi} L.,  {Salasnich} B.,  {Dal Cero} C.,
  {Rubele} S.,   {Nanni} A.,  2012, \mn@doi [\mnras]
  {10.1111/j.1365-2966.2012.21948.x}, 427, 127

\bibitem[\protect\citeauthoryear{{Carballo-Bello}}{{Carballo-Bello}}{2019}]{carballobello2019}
{Carballo-Bello} J.~A.,  2019, \mn@doi [\mnras] {10.1093/mnras/stz962}, \href
  {https://ui.adsabs.harvard.edu/abs/2019MNRAS.486.1667C} {486, 1667}

\bibitem[\protect\citeauthoryear{{Carpintero}, {G{\'o}mez}  \&
  {Piatti}}{{Carpintero} et~al.}{2013}]{carpinteroetal2013}
{Carpintero} D.~D.,  {G{\'o}mez} F.~A.,   {Piatti} A.~E.,  2013, \mn@doi
  [\mnras] {10.1093/mnrasl/slt096}, \href
  {http://adsabs.harvard.edu/abs/2013MNRAS.435L..63C} {435, L63}

\bibitem[\protect\citeauthoryear{{Crowl}, {Sarajedini}, {Piatti}, {Geisler},
  {Bica}, {Clari{\'a}}  \& {Santos}}{{Crowl} et~al.}{2001}]{cetal01}
{Crowl} H.~H.,  {Sarajedini} A.,  {Piatti} A.~E.,  {Geisler} D.,  {Bica} E.,
  {Clari{\'a}} J.~J.,   {Santos} Jr. J.~F.~C.,  2001, \mn@doi [\aj]
  {10.1086/321128}, 122, 220

\bibitem[\protect\citeauthoryear{{De Leo}, {Carrera}, {No{\"e}l}, {Read},
  {Erkal}  \& {Gallart}}{{De Leo} et~al.}{2020}]{deleoetal2020}
{De Leo} M.,  {Carrera} R.,  {No{\"e}l} N. E.~D.,  {Read} J.~I.,  {Erkal} D.,
  {Gallart} C.,  2020, \mn@doi [\mnras] {10.1093/mnras/staa1122}, \href
  {https://ui.adsabs.harvard.edu/abs/2020MNRAS.495...98D} {495, 98}

\bibitem[\protect\citeauthoryear{{Di Teodoro} et~al.,}{{Di Teodoro}
  et~al.}{2019}]{diteodoroetal2019}
{Di Teodoro} E.~M.,  et~al., 2019, \mn@doi [\mnras] {10.1093/mnras/sty3095},
  \href {http://adsabs.harvard.edu/abs/2019MNRAS.483..392D} {483, 392}

\bibitem[\protect\citeauthoryear{{Dias}, {Kerber}, {Barbuy}, {Bica}  \&
  {Ortolani}}{{Dias} et~al.}{2016}]{diasetal2016}
{Dias} B.,  {Kerber} L.,  {Barbuy} B.,  {Bica} E.,   {Ortolani} S.,  2016,
  \mn@doi [\aap] {10.1051/0004-6361/201527558}, 591, A11

\bibitem[\protect\citeauthoryear{{Dias} et~al.,}{{Dias}
  et~al.}{2021}]{diasetal2021}
{Dias} B.,  et~al., 2021, \mn@doi [\aap] {10.1051/0004-6361/202040015}, \href
  {https://ui.adsabs.harvard.edu/abs/2021A&A...647L...9D} {647, L9}

\bibitem[\protect\citeauthoryear{{Diaz} \& {Bekki}}{{Diaz} \&
  {Bekki}}{2012}]{db2012}
{Diaz} J.~D.,  {Bekki} K.,  2012, \mn@doi [\apj] {10.1088/0004-637X/750/1/36},
  \href {https://ui.adsabs.harvard.edu/abs/2012ApJ...750...36D} {750, 36}

\bibitem[\protect\citeauthoryear{{Dotter}, {Chaboyer}, {Ferguson}, {Lee},
  {Worthey}, {Jevremovi{\'c}}  \& {Baron}}{{Dotter}
  et~al.}{2007}]{dotteretal2007}
{Dotter} A.,  {Chaboyer} B.,  {Ferguson} J.~W.,  {Lee} H.-c.,  {Worthey} G.,
  {Jevremovi{\'c}} D.,   {Baron} E.,  2007, \mn@doi [\apj] {10.1086/519946},
  \href {https://ui.adsabs.harvard.edu/abs/2007ApJ...666..403D} {666, 403}

\bibitem[\protect\citeauthoryear{{Gaia Collaboration} et~al.,}{{Gaia
  Collaboration} et~al.}{2020}]{gaiaetal2020b}
{Gaia Collaboration} et~al., 2020, arXiv e-prints, \href
  {https://ui.adsabs.harvard.edu/abs/2020arXiv201201771G} {p. arXiv:2012.01771}

\bibitem[\protect\citeauthoryear{{Glatt} et~al.,}{{Glatt}
  et~al.}{2008}]{glattetal2008a}
{Glatt} K.,  et~al., 2008, \mn@doi [\aj] {10.1088/0004-6256/136/4/1703}, \href
  {http://adsabs.harvard.edu/abs/2008AJ....136.1703G} {136, 1703}

\bibitem[\protect\citeauthoryear{{Graczyk} et~al.,}{{Graczyk}
  et~al.}{2020}]{graczyketal2020}
{Graczyk} D.,  et~al., 2020, arXiv e-prints, \href
  {https://ui.adsabs.harvard.edu/abs/2020arXiv201008754G} {p. arXiv:2010.08754}

\bibitem[\protect\citeauthoryear{{Grillmair} \& {Dionatos}}{{Grillmair} \&
  {Dionatos}}{2006a}]{gd2006a}
{Grillmair} C.~J.,  {Dionatos} O.,  2006a, \mn@doi [\apjl] {10.1086/503744},
  \href {https://ui.adsabs.harvard.edu/abs/2006ApJ...641L..37G} {641, L37}

\bibitem[\protect\citeauthoryear{{Grillmair} \& {Dionatos}}{{Grillmair} \&
  {Dionatos}}{2006b}]{gd2006b}
{Grillmair} C.~J.,  {Dionatos} O.,  2006b, \mn@doi [\apjl] {10.1086/505111},
  \href {https://ui.adsabs.harvard.edu/abs/2006ApJ...643L..17G} {643, L17}

\bibitem[\protect\citeauthoryear{{Jacyszyn-Dobrzeniecka}
  et~al.,}{{Jacyszyn-Dobrzeniecka} et~al.}{2020}]{jacyszyndobrzenieckaetal2020}
{Jacyszyn-Dobrzeniecka} A.~M.,  et~al., 2020, \mn@doi [\apj]
  {10.3847/1538-4357/ab61f2}, \href
  {https://ui.adsabs.harvard.edu/abs/2020ApJ...889...26J} {889, 26}

\bibitem[\protect\citeauthoryear{{Koposov} et~al.,}{{Koposov}
  et~al.}{2018}]{koposovetal2018}
{Koposov} S.~E.,  et~al., 2018, \mn@doi [\mnras] {10.1093/mnras/sty1772}, \href
  {http://adsabs.harvard.edu/abs/2018MNRAS.479.5343K} {479, 5343}

\bibitem[\protect\citeauthoryear{{Kuzma}, {Da Costa}, {Mackey}  \&
  {Roderick}}{{Kuzma} et~al.}{2016}]{kuzmaetal2016}
{Kuzma} P.~B.,  {Da Costa} G.~S.,  {Mackey} A.~D.,   {Roderick} T.~A.,  2016,
  \mn@doi [\mnras] {10.1093/mnras/stw1561}, \href
  {http://adsabs.harvard.edu/abs/2016MNRAS.461.3639K} {461, 3639}

\bibitem[\protect\citeauthoryear{{Mackey}, {Koposov}, {Da Costa}, {Belokurov},
  {Erkal}  \& {Kuzma}}{{Mackey} et~al.}{2018}]{mackeyetal2018}
{Mackey} D.,  {Koposov} S.,  {Da Costa} G.,  {Belokurov} V.,  {Erkal} D.,
  {Kuzma} P.,  2018, \mn@doi [\apjl] {10.3847/2041-8213/aac175}, \href
  {http://adsabs.harvard.edu/abs/2018ApJ...858L..21M} {858, L21}

\bibitem[\protect\citeauthoryear{{Mart{\'\i}nez-V{\'a}zquez}, {Salinas}  \&
  {Vivas}}{{Mart{\'\i}nez-V{\'a}zquez} et~al.}{2021}]{mvetal2021}
{Mart{\'\i}nez-V{\'a}zquez} C.~E.,  {Salinas} R.,   {Vivas} A.~K.,  2021,
  \mn@doi [\aj] {10.3847/1538-3881/abd55e}, \href
  {https://ui.adsabs.harvard.edu/abs/2021AJ....161..120M} {161, 120}

\bibitem[\protect\citeauthoryear{{Niederhofer} et~al.,}{{Niederhofer}
  et~al.}{2021}]{niederhoferetal2021}
{Niederhofer} F.,  et~al., 2021, \mn@doi [\mnras] {10.1093/mnras/stab206},
  \href {https://ui.adsabs.harvard.edu/abs/2021MNRAS.tmp..268N} {}

\bibitem[\protect\citeauthoryear{{Olszewski}, {Saha}, {Knezek}, {Subramaniam},
  {de Boer}  \& {Seitzer}}{{Olszewski} et~al.}{2009}]{olszewskietal2009}
{Olszewski} E.~W.,  {Saha} A.,  {Knezek} P.,  {Subramaniam} A.,  {de Boer} T.,
   {Seitzer} P.,  2009, \mn@doi [\aj] {10.1088/0004-6256/138/6/1570}, \href
  {http://adsabs.harvard.edu/abs/2009AJ....138.1570O} {138, 1570}

\bibitem[\protect\citeauthoryear{{Omkumar} et~al.,}{{Omkumar}
  et~al.}{2021}]{omkumaretal2021}
{Omkumar} A.~O.,  et~al., 2021, \mn@doi [\mnras] {10.1093/mnras/staa3085},
  \href {https://ui.adsabs.harvard.edu/abs/2021MNRAS.500.2757O} {500, 2757}

\bibitem[\protect\citeauthoryear{{Parisi} et~al.,}{{Parisi}
  et~al.}{2014}]{paetal14}
{Parisi} M.~C.,  et~al., 2014, \mn@doi [\aj] {10.1088/0004-6256/147/4/71}, 147,
  71

\bibitem[\protect\citeauthoryear{{Piatti}}{{Piatti}}{2011}]{p11b}
{Piatti} A.~E.,  2011, \mn@doi [\mnras] {10.1111/j.1745-3933.2011.01145.x},
  418, L69

\bibitem[\protect\citeauthoryear{{Piatti}}{{Piatti}}{2012}]{p12b}
{Piatti} A.~E.,  2012, \mn@doi [\apjl] {10.1088/2041-8205/756/2/L32}, 756, L32

\bibitem[\protect\citeauthoryear{{Piatti}}{{Piatti}}{2014}]{p14c}
{Piatti} A.~E.,  2014, \mn@doi [\mnras] {10.1093/mnras/stu1917}, 445, 2302

\bibitem[\protect\citeauthoryear{{Piatti}}{{Piatti}}{2021a}]{piatti2021b}
{Piatti} A.~E.,  2021a, arXiv e-prints, \href
  {https://ui.adsabs.harvard.edu/abs/2021arXiv210403750P} {p. arXiv:2104.03750}

\bibitem[\protect\citeauthoryear{{Piatti}}{{Piatti}}{2021b}]{piatti2021a}
{Piatti} A.~E.,  2021b, \mn@doi [\aap] {10.1051/0004-6361/202039888}, \href
  {https://ui.adsabs.harvard.edu/abs/2021A&A...647A..11P} {647, A11}

\bibitem[\protect\citeauthoryear{{Piatti}, {Ivanov}, {Rubele}, {Marconi},
  {Ripepi}, {Cioni}, {Oliveira}  \& {Bekki}}{{Piatti} et~al.}{2016}]{petal2016}
{Piatti} A.~E.,  {Ivanov} V.~D.,  {Rubele} S.,  {Marconi} M.,  {Ripepi} V.,
  {Cioni} M.-R.~L.,  {Oliveira} J.~M.,   {Bekki} K.,  2016, \mn@doi [\mnras]
  {10.1093/mnras/stw1000}, \href
  {http://adsabs.harvard.edu/abs/2016MNRAS.460..383P} {460, 383}

\bibitem[\protect\citeauthoryear{{Piatti}, {Salinas}  \& {Grebel}}{{Piatti}
  et~al.}{2018}]{piattietal2018a}
{Piatti} A.~E.,  {Salinas} R.,   {Grebel} E.~K.,  2018, \mn@doi [\mnras]
  {10.1093/mnras/sty2761}, \href
  {http://adsabs.harvard.edu/abs/2018MNRAS.tmp.2632P} {}

\bibitem[\protect\citeauthoryear{{Piatti}, {Alfaro}  \&
  {Cantat-Gaudin}}{{Piatti} et~al.}{2019}]{piattietal2019}
{Piatti} A.~E.,  {Alfaro} E.~J.,   {Cantat-Gaudin} T.,  2019, \mn@doi [\mnras]
  {10.1093/mnrasl/sly240}, \href
  {https://ui.adsabs.harvard.edu/abs/2019MNRAS.484L..19P} {484, L19}

\bibitem[\protect\citeauthoryear{{Pieres} et~al.,}{{Pieres}
  et~al.}{2017}]{pieresetal2017}
{Pieres} A.,  et~al., 2017, \mn@doi [\mnras] {10.1093/mnras/stx507}, 468, 1349

\bibitem[\protect\citeauthoryear{{Pietrinferni}, {Cassisi}, {Salaris}  \&
  {Castelli}}{{Pietrinferni} et~al.}{2004}]{pietrinfernietal2004}
{Pietrinferni} A.,  {Cassisi} S.,  {Salaris} M.,   {Castelli} F.,  2004,
  \mn@doi [\apj] {10.1086/422498}, \href
  {https://ui.adsabs.harvard.edu/abs/2004ApJ...612..168P} {612, 168}

\bibitem[\protect\citeauthoryear{{Price-Whelan}, {Mateu}, {Iorio}, {Pearson},
  {Bonaca}  \& {Belokurov}}{{Price-Whelan} et~al.}{2019}]{pricewhelanetal2019}
{Price-Whelan} A.~M.,  {Mateu} C.,  {Iorio} G.,  {Pearson} S.,  {Bonaca} A.,
  {Belokurov} V.,  2019, \mn@doi [\aj] {10.3847/1538-3881/ab4cef}, \href
  {https://ui.adsabs.harvard.edu/abs/2019AJ....158..223P} {158, 223}

\bibitem[\protect\citeauthoryear{{Ripepi} et~al.,}{{Ripepi}
  et~al.}{2017}]{ripepietal2017}
{Ripepi} V.,  et~al., 2017, \mn@doi [\mnras] {10.1093/mnras/stx2096}, \href
  {https://ui.adsabs.harvard.edu/abs/2017MNRAS.472..808R} {472, 808}

\bibitem[\protect\citeauthoryear{{Shipp}, {Price-Whelan}, {Tavangar}, {Mateu}
  \& {Drlica-Wagner}}{{Shipp} et~al.}{2020}]{shippetal2020}
{Shipp} N.,  {Price-Whelan} A.~M.,  {Tavangar} K.,  {Mateu} C.,
  {Drlica-Wagner} A.,  2020, \mn@doi [\aj] {10.3847/1538-3881/abbd3a}, \href
  {https://ui.adsabs.harvard.edu/abs/2020AJ....160..244S} {160, 244}

\bibitem[\protect\citeauthoryear{{Sollima}}{{Sollima}}{2020}]{sollima2020}
{Sollima} A.,  2020, \mn@doi [\mnras] {10.1093/mnras/staa1209}, \href
  {https://ui.adsabs.harvard.edu/abs/2020MNRAS.495.2222S} {495, 2222}

\bibitem[\protect\citeauthoryear{{Wagner-Kaiser} \&
  {Sarajedini}}{{Wagner-Kaiser} \& {Sarajedini}}{2017}]{wagnerkaiser17}
{Wagner-Kaiser} R.,  {Sarajedini} A.,  2017, \mn@doi [\mnras]
  {10.1093/mnras/stw3206}, \href
  {http://adsabs.harvard.edu/abs/2017MNRAS.466.4138W} {466, 4138}

\bibitem[\protect\citeauthoryear{{Zivick} et~al.,}{{Zivick}
  et~al.}{2018}]{zivicketal2018}
{Zivick} P.,  et~al., 2018, \mn@doi [\apj] {10.3847/1538-4357/aad4b0}, \href
  {https://ui.adsabs.harvard.edu/abs/2018ApJ...864...55Z} {864, 55}

\bibitem[\protect\citeauthoryear{{Zivick}, {Kallivayalil}  \& {van der
  Marel}}{{Zivick} et~al.}{2020}]{zivicketal2020}
{Zivick} P.,  {Kallivayalil} N.,   {van der Marel} R.~P.,  2020, arXiv
  e-prints, \href {https://ui.adsabs.harvard.edu/abs/2020arXiv201102525Z} {p.
  arXiv:2011.02525}

\bibitem[\protect\citeauthoryear{{de Vaucouleurs} \& {Freeman}}{{de
  Vaucouleurs} \& {Freeman}}{1972}]{dVF1972}
{de Vaucouleurs} G.,  {Freeman} K.~C.,  1972, \mn@doi [Vistas in Astronomy]
  {10.1016/0083-6656(72)90026-8}, \href
  {https://ui.adsabs.harvard.edu/abs/1972VA.....14..163D} {14, 163}

\bibitem[\protect\citeauthoryear{{van der Marel}, {Alves}, {Hardy}  \&
  {Suntzeff}}{{van der Marel} et~al.}{2002}]{vdmareletal2002}
{van der Marel} R.~P.,  {Alves} D.~R.,  {Hardy} E.,   {Suntzeff} N.~B.,  2002,
  \mn@doi [\aj] {10.1086/343775}, \href
  {http://adsabs.harvard.edu/abs/2002AJ....124.2639V} {124, 2639}

\makeatother
\end{thebibliography}







\bsp	
\label{lastpage}
\end{document}